\documentclass[
reprint,
amsmath,amssymb,
aps,
pre,
floatfix,
]{revtex4-2}

\usepackage{graphicx}
\usepackage{bm}
\usepackage{color}
\usepackage{ulem}

\begin{document}

\title{Irreversibility of many-body harmonic oscillators characterized by generalized Husimi's adiabaticity parameter}

\author{Kyosuke Watanabe}
\thanks{Contact author: 9604830029@g.ecc.u-tokyo.ac.jp}
\affiliation{
Department of Complexity Science and Engineering,
Graduate School of Frontier Sciences,
The University of Tokyo,
Kashiwa, 277-8561, Japan
}

\author{Yuki Izumida}
\affiliation{
Department of Complexity Science and Engineering,
Graduate School of Frontier Sciences,
The University of Tokyo,
Kashiwa, 277-8561, Japan
}

\begin{abstract}
In a classical paper [K. Husimi, Prog. Theor. Phys. \textbf{9}, 381
(1953)], Husimi showed that, for a single harmonic oscillator with a
time-dependent angular frequency and initial conditions sampled from an equilibrium distribution, the average energy after a cyclic operation cannot be lower than its initial value. This irreversibility is quantified by
Husimi's adiabaticity parameter defined in terms of adiabatic invariants. In this work, we generalize Husimi's
framework to a many-body system of one-dimensional harmonic
oscillators coupled on an arbitrary connected network, with all spring
constants sharing a common time dependence. For a system attached to
a fixed wall, the dynamics can be decomposed into independent normal
modes by transforming to mass-weighted coordinates and diagonalizing
the resulting positive definite matrix, with each mode characterized
by its own Husimi's
adiabaticity parameter. By defining the generalized Husimi's
adiabaticity parameter as their arithmetic mean, we derive the exact
time evolution of the average energy and establish its
non-decrease under cyclic operations. Numerical simulations confirm these results for both a uniform nearest-neighbor chain and a heterogeneous network.
\end{abstract}

\maketitle

\section{Introduction}
In modern nonequilibrium statistical mechanics, 
the second law of thermodynamics can be derived from nonequilibrium equalities obtained by combining deterministic or stochastic dynamical evolution with initial equilibrium distributions~\cite{Jarzynski1997,Crooks1999,Seifert2005,Kawai2007,Jarzynski2011,Seifert2012}.
A representative example is the Jarzynski equality, which relates the
work performed during an arbitrary nonequilibrium process to the
equilibrium free-energy difference between the initial and final
parameter values~\cite{Jarzynski1997}.
From the Jarzynski equality, one can derive the maximum work principle:
\begin{equation}
    \langle W\rangle \ge \Delta F,
\end{equation}
where $\langle \cdot \rangle$ denotes an ensemble average, $\langle W\rangle$ denotes the averaged work performed on the system, and $\Delta F$ denotes the Helmholtz free-energy difference.
For a cyclic protocol, in which the externally controlled parameters
return to their initial values, $\Delta F=0$, and the Kelvin principle follows:
\begin{equation}\label{eq:W_ineq}
    \langle W\rangle \ge 0.
\end{equation}
These results apply not only to thermally isolated Hamiltonian dynamics~\cite{Jarzynski1997}, but also to stochastic dynamics in contact with a heat bath~\cite{Jarzynski1997_2,Crooks1999} and to certain deterministic thermostatted dynamics~\cite{Jarzynski1997,Jarzynski1997_2,SchollPaschinger2006}.

In the present work, we focus on the thermally isolated case described by a Hamiltonian. 
To be more specific, consider a Hamiltonian system initially prepared in the canonical distribution at inverse temperature $\beta$ and subsequently
isolated from the heat bath. Its Hamiltonian
$H(\bm{\alpha}(t))$ depends on externally controllable,
time-dependent parameters $\bm{\alpha}(t)$, which are varied from
$t=0$ to $t=\tau$.
Because the system is thermally isolated during the operation, its energy
change is entirely due to the work performed on the system by the external agent,
so that $W = H_\tau-H_0$ holds.
For a cyclic operation satisfying
$\bm{\alpha}(0)=\bm{\alpha}(\tau)$, the inequality
$\langle W\rangle\ge0$ in Eq.~(\ref{eq:W_ineq}) gives
\begin{equation}
    \label{eq:passivity}
    \langle H_\tau\rangle
    \ge
    \langle H_0\rangle,
\end{equation}
which can be interpreted as the Planck principle.

Although Eq.~(\ref{eq:passivity}) holds for arbitrary thermally
isolated Hamilton systems initially prepared in the canonical distribution under cyclic operation, it constrains only the sign of the average energy
change. It does not by itself provide an explicit dynamical quantity
that quantifies the excess energy generated by the finite-time operation.

A complementary perspective is provided by Husimi's analysis of a
single harmonic oscillator with a time-dependent angular frequency~\cite{Husimi1953}. Husimi showed that the average energy at time $t$
can be expressed exactly in terms of an adiabaticity parameter
$Q_t^{*}$ as
\begin{equation}
    \langle H_t\rangle
    =
    \langle H_0\rangle
    \frac{\omega_t}{\omega_0}
    Q_t^{*}.
\end{equation}
Here, $\omega_t$ denotes the angular frequency. The adiabaticity parameter $Q_t^*$ is defined in terms of adiabatic invariants in Eq.~(\ref{eq:definition_of_Husimi's_parameter}) and satisfies $Q_t^*\ge1$.
For a cyclic protocol satisfying $\omega_\tau=\omega_0$, this relation
reduces to
\begin{equation}
    \langle H_\tau\rangle
    =
    \langle H_0\rangle Q_\tau^{*}
    \ge
    \langle H_0\rangle,
\end{equation}
thus reproducing the inequality in Eq.~(\ref{eq:passivity})
through an explicit dynamical quantity.

Husimi's adiabaticity parameter characterizes the departure from
adiabatic dynamics: $Q_t^{*}=1$ is attained in the adiabatic limit reflecting the conservation of adiabatic invariants,
whereas $Q_t^{*}>1$ quantifies nonadiabatic excitation.
Owing to its analytical tractability, Husimi's adiabaticity parameter has been widely used to characterize irreversibility in classical~\cite{Sone2021,Izumida2025} and quantum~\cite{Deffner2008,Jaramillo2016,Jaramillo2017,Abah2012,Rossnagel2014,Beau2016,Mishima2017,Abah2018,Abah2019,Soriani2022,Abah_2014,PhysRevE.92.042148} time-dependent harmonic oscillators.
Recently, in~\cite{Izumida2025}, one of the authors extended Husimi's framework to a damped
time-dependent harmonic oscillator and clarified the relation between
the adiabaticity parameter and second-law-like inequalities for both cyclic and noncyclic processes.

These developments demonstrate that Husimi's adiabaticity parameter
provides an explicit dynamical characterization of irreversibility for
the time-dependent harmonic oscillator. Nevertheless, it remains unclear how this parameter can be generalized to interacting many-body systems and whether the resulting quantity can explicitly characterize their energy increase under cyclic operation.

In this paper, we develop such a formulation for a system of coupled
harmonic oscillators on an arbitrary connected network, whose
spring constants are modulated by a common time-dependent driving
protocol. By exploiting the spectral decomposition of the network
matrix, we decompose the coupled harmonic-oscillators dynamics into independent normal-modes dynamics, where each normal mode is described as a time-dependent harmonic
oscillator. We associate Husimi's adiabaticity parameter with each
normal mode and introduce the generalized Husimi's adiabaticity parameter
as their arithmetic mean. 
We then show that the total energy under
cyclic operation can be expressed directly in terms of this generalized
parameter, providing an explicit mechanical derivation of Eq.~(\ref{eq:passivity})
for the many-body system.

The remainder of this paper is organized as follows.
In Sec.~\ref{sec:Husimi's_work}, we review Husimi's original result for
a single harmonic oscillator.
In Sec.~\ref{sec:model}, we introduce many-body harmonic
oscillators on a network and present its normal-mode decomposition.
In Sec.~\ref{sec:main_results}, we introduce the generalized
Husimi's adiabaticity parameter and show how the inequality
Eq.~(\ref{eq:passivity}) follows from this parameter in the many-body
case.
In Sec.~\ref{sec:numerical_results}, we verify our theoretical results
by numerical simulations.
In Sec.~\ref{sec:discussion}, we discuss the limitations of the present
formulation.
Finally, in Sec.~\ref{sec:concluding_remarks},
concluding remarks are given.

\section{Husimi's result for a single harmonic oscillator}\label{sec:Husimi's_work}

For the sake of completeness, we review the original result for a single harmonic oscillator by Husimi~\cite{Husimi1953} (see also Ref.~\cite{Izumida2025}).
We consider a one-dimensional harmonic oscillator with unit mass and a time-dependent angular frequency $\omega_t > 0$, obeying the following equation of motion:
\begin{equation} \label{eq:EOM_of_one_oscillator}
    \ddot{x}_t = -\omega_t^2 x_t.
\end{equation}
Defining the momentum by $p_t \equiv \dot{x}_t$, the Hamiltonian is given by
\begin{equation}
    H_t(x_t,p_t)
    =
    \frac{p_t^2}{2}
    +
    \frac{\omega_t^2 x_t^2}{2}.
\end{equation}
Here, the quantity
\begin{equation} \label{eq:definition_of_adiabatic_invariant}
    \frac{H_t(x_t,p_t)}{\omega_t}
    =
    \frac{
    p_t^2+\omega_t^2 x_t^2
    }{
    2\omega_t
    }
\end{equation}
is approximately conserved along a trajectory as an adiabatic invariant
when $\omega_t$ varies adiabatically (sufficiently slowly).

Husimi showed that, when the oscillator is initially sampled from an equilibrium distribution, its average energy can be written as
\begin{equation} \label{eq:E_t_in_terms_of_Q_t^*}
    E_t
    =
    E_0
    \,
    \frac{\omega_t}{\omega_0}
    \,
    Q_t^*,
\end{equation}
where $E_t \equiv \langle H_t \rangle$ is the average energy at time $t$ and $Q_t^*$ is the adiabaticity parameter introduced by Husimi~\cite{Husimi1953}, whose definition will be given in Eq.~(\ref{eq:definition_of_Husimi's_parameter}). Furthermore, it is shown that $Q_t^*$ satisfies the following inequality~\cite{Husimi1953,Izumida2025}
\begin{equation} \label{eq:Q_t^*_for_no_friction}
    Q_t^* \ge 1,
\end{equation}
where the equality is achieved when the angular frequency is varied adiabatically.
Therefore, for any cyclic operation satisfying
\begin{equation}
    \omega_\tau = \omega_0,
\end{equation}
Eq.~(\ref{eq:E_t_in_terms_of_Q_t^*}) reduces to
\begin{equation}
    E_\tau = E_0 Q_\tau^*,\label{eq.inq_E}
\end{equation}
which, combined with Eq.~(\ref{eq:Q_t^*_for_no_friction}), immediately yields
\begin{equation}
    E_\tau \ge E_0.
\end{equation}
Thus, for any cyclic operation returning the angular frequency to its initial value, the final average energy is no lower than the initial average energy. This may provide a mechanical derivation of irreversibility analogous to the Planck principle in thermodynamics~\cite{Izumida2025}.

We now review the derivation of Eq.~(\ref{eq:E_t_in_terms_of_Q_t^*}). Let $X_t$ and $Y_t$ be the fundamental solutions of Eq.~(\ref{eq:EOM_of_one_oscillator}) satisfying
\begin{equation} \label{eq:definition_of_X_t}
    \ddot{X}_t = -\omega_t^2 X_t,
    \qquad
    X_0 = 0,
    \qquad
    \dot{X}_0 = 1,
\end{equation}
and
\begin{equation} \label{eq:definition_of_Y_t}
    \ddot{Y}_t = -\omega_t^2 Y_t,
    \qquad
    Y_0 = 1,
    \qquad
    \dot{Y}_0 = 0.
\end{equation}
For an initial condition $(x_0,p_0)$, the time evolution of the oscillator can be expressed as
\begin{equation}
    x_t = x_0 Y_t + p_0 X_t,
\end{equation}
and
\begin{equation}
    p_t = x_0 \dot{Y}_t + p_0 \dot{X}_t
\end{equation}
in terms of the fundamental solutions $X_t$ and $Y_t$.

Assume that the oscillator initially obeys the canonical distribution with inverse temperature $\beta$:
\begin{equation}
    f_0(x_0,p_0)
    =
    \frac{e^{-\beta H_0(x_0,p_0)}}{Z},\label{eq:canonical}
\end{equation}
where $Z$ denotes the partition function as
\begin{equation}
    Z
    \equiv \int_{-\infty}^\infty \int_{-\infty}^\infty dx_0dp_0 \exp \left[ -\beta H_0(x_0,p_0) \right]
    =
    \frac{2\pi}{\beta\omega_0}.
\end{equation}
The initial average energy is therefore given by
\begin{equation}
    E_0
    =
    \langle H_0 \rangle
    =
    -\frac{\partial}{\partial \beta}\ln Z
    =
    \frac{1}{\beta}.
\end{equation}

Since the dynamics is deterministic, the probability is conserved along a trajectory:
\begin{equation}
    f_t(x_t,p_t)\,dx_t\,dp_t
    =
    f_0(x_0,p_0)\,dx_0\,dp_0.
\end{equation}
Using this relation, the average energy at time $t$ can be written as
\begin{align}
    E_t
    &=\int_{-\infty}^\infty \int_{-\infty}^\infty
    dx_t\,dp_t\,
    f_t(x_t,p_t)H_t(x_t,p_t)\nonumber\\
    &=
    \int_{-\infty}^\infty \int_{-\infty}^\infty
    dx_0\,dp_0\,
    \frac{e^{-\beta H_0(x_0,p_0)}}{Z}
    \,
    H_t\bigl(x_t(x_0,p_0),p_t(x_0,p_0)\bigr).
\end{align}
Performing this integration yields Eq.~(\ref{eq:E_t_in_terms_of_Q_t^*}), with Husimi's adiabaticity parameter defined as follows:
\begin{equation} \label{eq:definition_of_Husimi's_parameter}
    Q_t^*
    \equiv
    \omega_0
    \frac{
    \dot{X}_t^2 + \omega_t^2 X_t^2
    }{
    2\omega_t
    }
    +
    \frac{1}{\omega_0}
    \frac{
    \dot{Y}_t^2 + \omega_t^2 Y_t^2
    }{
    2\omega_t
    }.
\end{equation}
Husimi's adiabaticity parameter $Q_t^*$ is given by a linear combination of the adiabatic invariants of the fundamental solutions $X_t$ and $Y_t$ obeying Eqs.~(\ref{eq:definition_of_X_t}) and (\ref{eq:definition_of_Y_t}). Consequently, $Q_t^*=1$ is attained in the adiabatic limit, while rapid changes of $\omega_t$ generally lead to $Q_t^*>1$. Therefore, $Q_t^*$ quantitatively characterizes the degree of irreversibility associated with the operation of the angular frequency.

\section{Model}\label{sec:model}
\subsection{Network-coupled harmonic oscillators} \label{section:model_of_N_body_wall}
We consider a one-dimensional system of $N$ point masses. The $i$-th point mass ($i=1,2,\dots,N$) with mass $m_i$ has displacement $x_t^{(i)}$ from an equilibrium point, which can be regarded as a harmonic oscillator.
Any pair of point masses can be connected by a spring, and the total system forms a network of coupled harmonic oscillators.
The first point mass may also be attached to a fixed wall by a spring.  

The network structure formed by springs between point masses is given by the set $E$ of spring-connected pairs $(i,j)$:
\begin{equation}
    E \subset \{(i,j)\mid 1\le i<j\le N\}.
\end{equation}
We assume that the network is connected, so that every point mass is mechanically coupled, either directly or indirectly, to every other point mass through the spring network.

We also assume that the time evolution of a spring constant between $i$ and $j$ factorizes as
\begin{equation}
    k_{ij}(t) = a_{ij}\omega_t^2, 
\end{equation}
where $a_{ij}=a_{ji}\ge 0$ is a time-independent coupling coefficient
with dimensions of mass. We take $a_{ij}>0$ when point masses $i$ and
$j$ are connected by a spring and $a_{ij}=0$ otherwise. Thus,
\[
(i,j)\in E \quad \Longleftrightarrow \quad k_{ij}(t)>0 \ \text{for all } t,
\]
\[
(i,j)\notin E \quad \Longleftrightarrow \quad k_{ij}(t)= 0 \ \text{for all } t.
\]
The wall spring that is attached to the first point mass is given by $k_1(t)=b_1\omega_t^2$ with $b_1>0$.

The equation of motion for the $i$-th point mass reads
\begin{equation} \label{eq:EOM_of_x_i}
    m_i \ddot{x}_t^{(i)} = -\omega_t^2 \left\{
    \sum_{j(\neq i)} a_{ij} \left( x_t^{(i)} - x_t^{(j)} \right) + b_1\delta_{i,1} x_t^{(1)} 
    \right\}.
\end{equation}
We define an $N\times N$ symmetric matrix $L$ by
\begin{equation}
    (L)_{ii}=\sum_{j(\neq i)} a_{ij}, 
    \qquad 
    (L)_{ij} =
    -a_{ij} \quad(i \ne j). 
\end{equation}
Furthermore, let $\bm{x}_t =(x_t^{(1)},\dots,x_t^{(N)})^{\top}$ and
\begin{equation}\label{eq:matrix_M_B}
M^\alpha=\mathrm{diag}\!\left(m_1^\alpha,\dots,m_N^\alpha\right),\ B=\mathrm{diag}(b_1,0,\dots,0).
\end{equation}
Then, Eq.~(\ref{eq:EOM_of_x_i}) can be rewritten in terms of the vector notation:
\begin{equation} \label{eq:EOM_of_N_body_nonsymmetric}
    \ddot{\bm{x}}_t = - \omega_t^2 M^{-1}(L+B) \bm{x}_t.
\end{equation}
With $\bm{p}_t=(p_t^{(1)},p_t^{(2)},\dots,p_t^{(N)})^{\top}=M\dot{\bm{x}}_t$, the Hamiltonian of this system is given by
\begin{align} \label{eq:hamiltonian_coupled}
    H_t
    &=\frac{1}{2} \bm{p}_t^{\top} M^{-1}\bm{p}_t + \frac{\omega_t^2}{2} \bm{x}_t^{\top} (L+B) \bm{x}_t.
\end{align}

\subsection{Diagonalization and mode decomposition}
Since $M^{-1}(L+B)$ is not symmetric in general, we introduce
\begin{equation} \label{eq:definition_of_x'_and_p'}
    \bm{x}'_t=M^{\frac{1}{2}}\bm{x}_t,
    \qquad
    \bm{p}'_t =M^{-\frac{1}{2}}\bm{p}_t.
\end{equation}
Then, Eqs.~(\ref{eq:EOM_of_N_body_nonsymmetric}) and (\ref{eq:hamiltonian_coupled}) become
\begin{equation}\label{eq:dynamics_x}
    \ddot{\bm{x}}'_t=-\omega_t^2 A{\bm{x}}'_t,
\end{equation}
and
\begin{equation} \label{eq:hamiltonian_still_coupled}
    H_t 
    =\frac{1}{2} {\bm{p}'_t}^{\top} \bm{p}'_t+\frac{\omega_t^2}{2}{\bm{x}'_t}^{\top} A\bm{x}'_t,
\end{equation}
where
\begin{equation} \label{eq:definition_of_A}
    A\equiv M^{-\frac{1}{2}}(L+B)M^{-\frac{1}{2}}.
\end{equation}
We refer to $A$ as the mass-weighted coupling matrix.
Since $A^{\top}=A$, $A$ is symmetric, and there exists an orthogonal matrix $S$ such that
\begin{equation}
    S^{\top} A S = \mathrm{diag}(\lambda_1,\dots, \lambda_N).
\end{equation}
We can show that all eigenvalues satisfy $\lambda_i>0$ for a connected network with the wall-attached case ($b_1>0$), see Appendix~\ref{subsec:proof_of_positive_definite}.

Normal coordinates are defined in terms of $S$:
\begin{equation}\label{eq:normal-mode_tarnsformation}
\bm{u}_t =\begin{pmatrix}
u_t^{(1)} \\
u_t^{(2)} \\
\vdots \\
u_t^{(N)}
\end{pmatrix}
=S^{\top} \bm{x}'_t, \qquad 
\bm{v}_t =\begin{pmatrix}
v_t^{(1)} \\
v_t^{(2)}  \\
\vdots \\
v_t^{(N)} 
\end{pmatrix}
=S^{\top} \bm{p}'_t.
\end{equation}
Then, Eq.~(\ref{eq:dynamics_x}) is decomposed into $N$ independent mode equations:
\begin{equation} \label{eq:EOM_of_Q_t^(i)}
    \ddot{u}_t^{(i)} = - \lambda_i \omega_t^2 u_t^{(i)} \quad (i=1,2,\dots ,N).
\end{equation}
Each mode is equivalent to a single harmonic oscillator with time-dependent angular frequency $\sqrt{\lambda_i}\omega_t$. Moreover, Eq.~(\ref{eq:hamiltonian_still_coupled}) becomes a sum of independent mode Hamiltonians:
\begin{equation}\label{eq:decomposition_mode_H}
    H_t 
    = \sum_{i=1}^{N}  \left[ \frac{1}{2} \left(v_t^{(i)}\right)^2+ \frac{\lambda_i \omega_t^2}{2} \left(u_t^{(i)}\right)^2 \right]
    \equiv \sum_{i=1}^{N} H_t^{(i)}.
\end{equation}

\section{Main Results}\label{sec:main_results}

In this Section, we introduce the generalized Husimi's adiabaticity parameter as the arithmetic mean of the mode-wise adiabaticity parameters and show that it characterizes the irreversibility of the many-body harmonic oscillators.

Assume that the initial distribution of $(\bm{x}_0,\bm{p}_0)$ obeys the canonical distribution with inverse temperature $\beta$. The partition function is given by
\begin{align}
Z_{\mathrm{w}}
&\equiv
\int \int d\bm{x}_0\, d\bm{p}_0
\exp \left[ -\beta H_0(\bm{x}_0,\bm{p}_0) \right]\nonumber\\ 
&=\int \int d\bm{u}_0\, d\bm{v}_0
\exp \left[ -\beta H_0(\bm{u}_0,\bm{v}_0) \right] \nonumber \\
&=
\frac{1}{\sqrt{\lambda_1\lambda_2\dots\lambda_N}}
\left(
\frac{2\pi}{\omega_0 \beta}
\right)^N,
\end{align}
where we used that the Jacobian of the transformation from $(\bm{x}_0, \bm{p}_0)$ to $(\bm{u}_0, \bm{v}_0)$ is unity in the second equality.
Thus, the initial average energy becomes
\begin{equation}
    E_0
    =
    \langle H_0 \rangle
    =
    -\frac{\partial}{\partial \beta}\ln Z_{\mathrm{w}}
    =
    \frac{N}{\beta}.
\end{equation}

As in the single-oscillator case, let $X_t^{(i)}$ and $Y_t^{(i)}$
be two fundamental solutions of Eq.~(\ref{eq:EOM_of_Q_t^(i)})
satisfying
\begin{equation}\label{eq:eq_of_fundamental_solution_X}
    \ddot{X}_t^{(i)} = - \lambda_i \omega_t^2 X_t^{(i)}, \quad X_0^{(i)}=0,\quad \dot{X}_0^{(i)}=1,
\end{equation}
and
\begin{equation}\label{eq:eq_of_fundamental_solution_Y}
    \ddot{Y}_t^{(i)} = - \lambda_i \omega_t^2 Y_t^{(i)}, \quad Y_0^{(i)}=1,\quad \dot{Y}_0^{(i)}=0.
\end{equation}
For an initial condition $(u_0^{(i)},v_0^{(i)})$, the solution
for each mode can then be written as
\begin{equation} \label{eq:ut_time_evolution}
    u_t^{(i)}
    =
    u_0^{(i)} Y_t^{(i)}
    +
    v_0^{(i)} X_t^{(i)},
\end{equation}
\begin{equation} \label{eq:vt_time_evolution}
    v_t^{(i)}
    =
    u_0^{(i)} \dot{Y}_t^{(i)}
    +
    v_0^{(i)} \dot{X}_t^{(i)}.
\end{equation}
A direct calculation yields (see Appendix~\ref{app:derivation_of_E_t} for the detailed derivation):
\begin{align} \label{eq:E_t_N-body}
    E_t
    =
    E_0
    \,
    \frac{\omega_t}{\omega_0}
    \,
    \overline{Q}_t^{*},
\end{align}
where we define the generalized Husimi's adiabaticity parameter by
\begin{equation} \label{eq:generalized_Husimi_parameter}
    \overline{Q}_t^{*}
    \equiv
    \frac{1}{N}
    \sum_{i=1}^{N}
    Q_t^{*(i)}.
\end{equation}
Here, the mode-wise adiabaticity parameter is
given by
\begin{align} \label{eq:Husimi's_parameter_N-body}
Q_t^{*(i)}
=
\sqrt{\lambda_i}\omega_0
&
\frac{
\left(\dot{X}_t^{(i)} \right)^2
+
\lambda_i \omega_t^2 \left(X_t^{(i)}\right)^2
}{
2\sqrt{\lambda_i}\omega_t
}
\nonumber \\
&+
\frac{1}{\sqrt{\lambda_i}\omega_0}
\frac{
\left(\dot{Y}_t^{(i)} \right)^2
+
\lambda_i \omega_t^2 \left(Y_t^{(i)}\right)^2
}{
2\sqrt{\lambda_i}\omega_t
}
\end{align}
for the $i$-th normal mode. Since each mode-wise adiabaticity parameter satisfies the following inequality (see Appendix~\ref{app:derivation_of_Q>1_N_body} for the derivation)
\begin{equation} \label{eq:inequality_for_Q_t^{(i)}}
    Q_t^{*(i)} \ge 1,
\end{equation}
the generalized adiabaticity parameter (\ref{eq:generalized_Husimi_parameter}) satisfies
\begin{equation}
    \overline{Q}_t^{*} \ge 1.
\end{equation}
Therefore, for a cyclic operation satisfying $\omega_\tau=\omega_0$, Eq.~(\ref{eq:E_t_N-body}) reduces to
\begin{equation} \label{eq:Planck_principle_N-body}
    E_\tau
    =
    E_0
    \overline{Q}_\tau^{*}
    \ge
    E_0.
\end{equation}
Thus, the irreversibility of the many-body harmonic oscillators can be characterized explicitly by the generalized Husimi's adiabaticity parameter.

Equation~(\ref{eq:E_t_N-body}), together with Eqs.~(\ref{eq:generalized_Husimi_parameter}) and (\ref{eq:Husimi's_parameter_N-body}), is the central result of this work. While the inequality $E_\tau \ge E_0$ itself may follow from more general arguments such as the Jarzynski equality~\cite{Jarzynski1997}, Eq.~(\ref{eq:E_t_N-body}) expresses the detailed form of the average energy in terms of the generalized Husimi's adiabaticity parameter, whose mode-resolved decomposition is given by Eq.~(\ref{eq:generalized_Husimi_parameter}). In particular, the network structure enters explicitly through the eigenvalues $\lambda_i$, which determine the contribution of each normal mode to the total irreversibility.
Furthermore, the extension of the present result to a system without the wall is possible. 
While all normal modes have positive frequencies in the present wall-attached case, translational zero mode appears in a case without the wall. See Appendix~\ref{app:no_wall_case} for the detailed formulation.

\section{Numerical demonstration}\label{sec:numerical_results}

\subsection{Method}
To verify that the generalized Husimi's adiabaticity parameter in Eq.~(\ref{eq:generalized_Husimi_parameter}) correctly characterizes the irreversibility of the many-body system, we performed numerical calculations for two network structures, a nearest-neighbor chain network and a heterogeneous network.

For Fig.~\ref{fig:theory_simulation_comparison}, the equations of motion (\ref{eq:EOM_of_N_body_nonsymmetric}) were solved numerically using the fourth-order Runge--Kutta method with a fixed time step $\Delta t=4\times10^{-4}$, and $10^4$ initial conditions for $(\bm{x}_0,\bm{p}_0)$ were sampled from the canonical distribution at inverse temperature $\beta$ using Gaussian random variables consistent with the initial Hamiltonian.
For Fig.~\ref{fig:tau_vs_energy_increase}, the equations for the fundamental solutions of each normal mode (\ref{eq:eq_of_fundamental_solution_X}) and (\ref{eq:eq_of_fundamental_solution_Y}) were solved using the same method, with the time step adjusted according to the protocol duration $\tau$. The time steps ranged approximately from $8\times10^{-6}$ to $2.6\times10^{-3}$ for the chain network and from $1.3\times10^{-5}$ to $7.1\times10^{-4}$ for the heterogeneous network.
Numerical convergence was examined by varying the number of sampled initial conditions for Fig.~\ref{fig:theory_simulation_comparison} and by halving the integration time step for both figures.

We use the following cyclic protocol of duration $\tau$:
\begin{equation}
    \omega_t
    =
    \omega_0
    +
    C
    \sin
    \left(
    \frac{2\pi t}{\tau}
    \right),
    \quad
    (0\le t\le\tau),
\end{equation}
where $C$ ($0<C<\omega_0$) is a constant.
Throughout this Section, we set $N=10$, $\beta=1$, $b_1=1$, $C=0.5$, and $\omega_0=1$.

We used ChatGPT (OpenAI, GPT-5.5) to generate the Python code used for the numerical simulations. The AI-generated code was carefully checked by the authors before use.

\subsection{Nearest-neighbor chain}

We first consider the simplest nearest-neighbor chain network with uniform masses and coupling constants. 
% (Fig.~\ref{fig:network_structure_chain}) 
The mass matrix is given by
\begin{equation*}
    M
    =
    \mathrm{diag}(1,1,\dots,1),
\end{equation*}
and the corresponding matrix $L$ is given by
\begin{equation}
L=
\begin{pmatrix}
1 & -1 & 0 & \cdots & 0 \\
-1 & 2 & -1 & \cdots & 0 \\
0 & -1 & 2 & \cdots & 0 \\
\vdots & \vdots & \vdots & \ddots & -1 \\
0 & 0 & 0 & -1 & 1
\end{pmatrix}.
\end{equation}

\begin{figure*}[t]
\centering
\includegraphics[width=\textwidth]{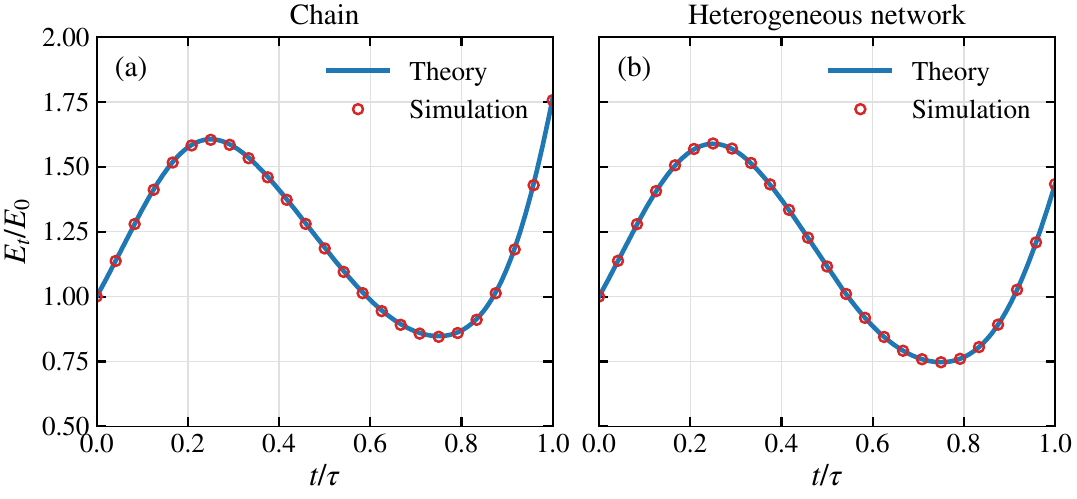}
\caption{Time evolution of the normalized average energy for (a) the nearest-neighbor chain and (b) the heterogeneous network. The blue solid curves represent the theoretical prediction of Eq.~(\ref{eq:E_t_N-body}), and the red open circles show direct numerical simulations averaged over $10^4$ initial conditions sampled from the canonical distribution.
We used $\tau=2$.}
\label{fig:theory_simulation_comparison}
\end{figure*}

Figure~\ref{fig:theory_simulation_comparison}(a)
shows the time evolution of the normalized average energy obtained by the direct numerical simulation (red open circles),
compared with the theory (\ref{eq:E_t_N-body}) using the generalized Husimi's adiabaticity parameter (\ref{eq:generalized_Husimi_parameter}) (blue solid curve).
The excellent agreement confirms that the evolution of the average energy is quantitatively characterized by the generalized Husimi's adiabaticity parameter. In particular, the increase in the average energy after the cyclic operation is consistent with the theoretical prediction (\ref{eq:Planck_principle_N-body}).

Next, we investigate the dependence of the degree of irreversibility on the protocol duration $\tau$. According to Eq.~(\ref{eq:Planck_principle_N-body}), the energy increase after one cycle is given by
\begin{equation} \label{eq:normalized_energy_increase_theory}
    \frac{\Delta E}{E_0}
    =
    \frac{E_\tau-E_0}{E_0}
    =
    \frac{1}{N}
    \sum_{i=1}^{N}
    \left[
    Q_\tau^{*(i)}-1
    \right].
\end{equation}

To characterize the intrinsic timescale of the system, we define
\begin{equation}
    T_{\mathrm{max}}
    \equiv
    \frac{2\pi}
    {\sqrt{\lambda_{\mathrm{min}}}\omega_0},
\end{equation}
where $\lambda_{\mathrm{min}}$ is the smallest eigenvalue of $A$. Thus, the ratio $\tau/T_{\mathrm{max}}$ compares the protocol duration with the slowest timescale of the system.

\begin{figure*}[t]
\centering
\includegraphics[width=\textwidth]{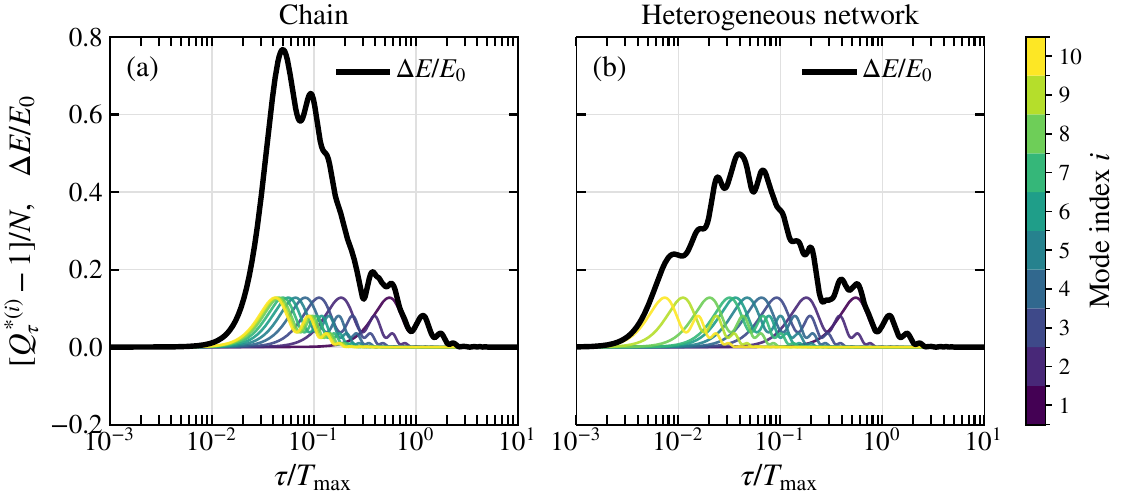}
\caption{Mode-resolved contributions $[Q_\tau^{*(i)}-1]/N$ (colored thin curves) and the total normalized energy increase $\Delta E/E_0$ (thick black curves) as functions of the scaled protocol duration $\tau/T_\mathrm{max}$ for (a) the nearest-neighbor chain with $T_\mathrm{max}=42.0$ and (b) the heterogeneous network with $T_\mathrm{max}=63.5$. At each protocol duration, the black curve is the sum of the colored mode contributions. The color bar indicates the mode index $i$, ordered by increasing eigenvalue $\lambda_i$.}
\label{fig:tau_vs_energy_increase}
\end{figure*}

Figure~\ref{fig:tau_vs_energy_increase}(a) shows both the total energy increase $\Delta E/E_0$ (black bold curve) and the mode-resolved contributions $[Q_\tau^{*(i)}-1]/N$ (colored thin curves) as functions of $\tau/T_{\mathrm{max}}$. 
As expected from Eq.~(\ref{eq:inequality_for_Q_t^{(i)}}),
all the mode contributions are nonnegative. They vanish in the adiabatic limit ($\tau/T_{\mathrm{max}}\to\infty$).
They also vanish in the sudden-driving limit ($\tau/T_{\mathrm{max}}\to0$) at fixed driving amplitude, because the mode coordinates and momenta remain essentially unchanged during the cycle, while the Hamiltonian returns to its initial form.

Each contribution $[Q_\tau^{*(i)}-1]/N$ develops a pronounced peak at an intermediate value of the protocol duration and subsequently decays toward zero as the dynamics approaches the adiabatic regime. The location of the peak depends strongly on the eigenvalue $\lambda_i$. Modes with larger eigenvalues reach their maxima at shorter protocol durations, whereas slower modes attain their maxima at longer protocol durations. Consequently, the peaks of different modes are distributed over a wide range of timescales.

As a result, the total energy increase is not a simple monotonic function of $\tau$. Instead, the multiple local maxima observed in $\Delta E/E_0$ originate from the superposition of the mode-resolved contributions. This behavior provides a direct illustration of how the irreversibility of the many-body system is built from the nonadiabatic excitations of individual normal modes.

\subsection{Heterogeneous network}
\begin{figure}[t!]
\centering
\includegraphics[width=\columnwidth]{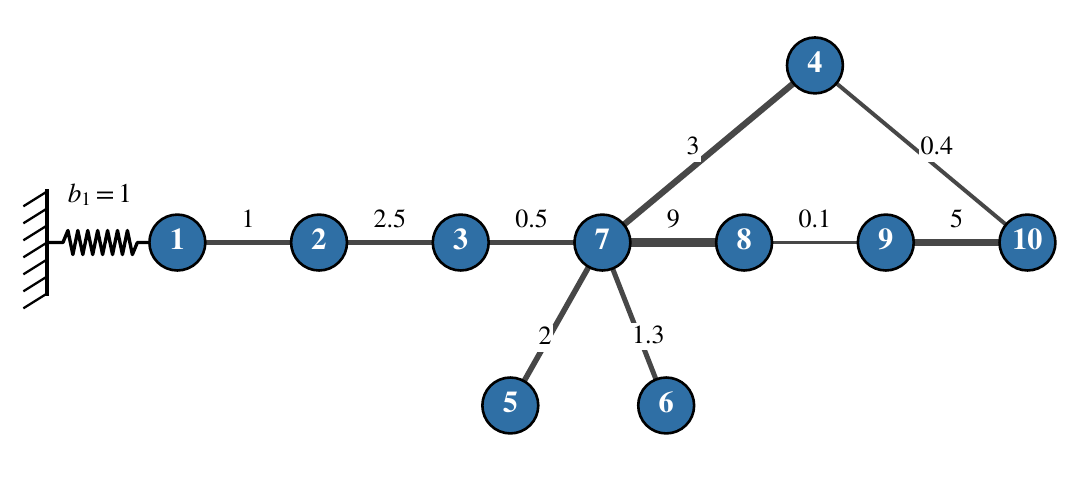}
\caption{Schematic of the heterogeneous network used in the numerical
simulations (Figs.~\ref{fig:theory_simulation_comparison}(b) and \ref{fig:tau_vs_energy_increase}(b)). The numbered circles denote the point masses, and the
numbers beside the edges give the coupling constants $a_{ij}$; the
edge widths also increase with coupling strength. Point mass 1 is
attached to a fixed wall by a spring with $b_1=1$. The nonuniform
masses are specified in Eq.~(\ref{eq:nonuniform_masses}).}
\label{fig:network_structure_general}
\end{figure}
To demonstrate that the theory is not restricted to uniform systems, we next consider a heterogeneous network with nonuniform masses and coupling constants (Fig.~\ref{fig:network_structure_general}). The mass matrix is chosen as
\begin{equation} \label{eq:nonuniform_masses}
    M
    =
    \mathrm{diag}
    \left(
    1,
    2,
    0.5,
    2.5,
    5,
    1,
    0.8,
    1.5,
    9,
    0.1
    \right).
\end{equation}

Even in this nonuniform system, the theoretical prediction for the average energy based on the generalized Husimi's adiabaticity parameter agrees quantitatively with the direct numerical simulations, as shown in Fig.~\ref{fig:theory_simulation_comparison}(b).

Figure~\ref{fig:tau_vs_energy_increase}(b) shows the dependence of the energy increase on the protocol duration $\tau$ for the heterogeneous network. Compared with the nearest-neighbor chain network, the heterogeneous network possesses a broader eigenvalue spectrum. Consequently, the maxima of the mode-resolved quantities $Q_\tau^{*(i)}-1$ are distributed over a wider range of protocol durations. Nevertheless, the same mechanism remains valid: the non-monotonic structure of the total energy increase originates from the superposition of mode-wise contributions.

\section{Discussion}\label{sec:discussion}
The present framework relies on the assumption that all spring
constants share the same time-dependent factor. Under this assumption,
the coefficient matrix in the mass-weighted equation of motion (\ref{eq:dynamics_x})
factorizes as $\omega_t^2 A$, where $A$ is time independent, and hence
the normal-mode transformation (\ref{eq:normal-mode_tarnsformation}) can be chosen independently of time.

If, instead, the spring constants vary independently, the equation of
motion in the mass-weighted coordinates generally takes the form
\begin{equation}
    \ddot{\bm{x}}'_t=-A_t\bm{x}'_t,
    \qquad
    A_t=M^{-1/2}(L_t+B_t)M^{-1/2},
\end{equation}
where $L_t$ is defined by
$(L_t)_{ii}=\sum_{j(\neq i)}k_{ij}(t)$ and
$(L_t)_{ij}=-k_{ij}(t)$ for $i\neq j$, and
$B_t=\mathrm{diag}(k_1(t),0,\dots,0)$.
Here, $k_{ij}(t)=0$ holds for pairs not connected by a spring.
Both the eigenvalues and eigenvectors of $A_t$ may depend on time. Let $S_t$ be an orthogonal matrix that instantaneously
diagonalizes $A_t$, and define $\bm{u}_t=S_t^{\top}\bm{x}'_t$.
The transformed equation of motion is then given by
\begin{equation} \label{eq:nondiagonal_term}
    \ddot{\bm{u}}_t
    +2S_t^{\top}\dot{S}_t\dot{\bm{u}}_t
    +S_t^{\top}\ddot{S}_t\bm{u}_t
    =
    -\left(S_t^{\top}A_tS_t\right)\bm{u}_t.
\end{equation}
Although $S_t^{\top}A_tS_t$ is diagonal, the terms involving
$\dot{S}_t$ and $\ddot{S}_t$ generally couple different instantaneous
normal modes. Consequently, the system cannot generally be decomposed
into dynamically independent oscillators, and the mode-wise
construction used in the present work is not directly applicable.

Nevertheless, even in such general systems, the
inequality $E_\tau \ge E_0$
still holds whenever the Hamiltonian returns to its initial form after the cycle, namely when
\begin{equation}
    k_{ij}(0)=k_{ij}(\tau)\quad ((i,j)\in E),
    \qquad
    k_1(0)=k_1(\tau).
\end{equation}
Therefore, what is broken in this general case is not irreversibility itself, but the possibility of expressing irreversibility as a sum of independent mode contributions.
An important open problem is whether one can define a generalized adiabaticity parameter that characterizes irreversibility in such mode-coupling and time-dependent systems.

\section{Concluding Remarks}\label{sec:concluding_remarks}
In this work, we derived an irreversibility relation for the average energy of many-body harmonic oscillators coupled on an arbitrary connected network under cyclic operations. By incorporating the eigenvalues of the positive definite matrix determined by the network structure and masses, we introduced the generalized Husimi's adiabaticity parameter as the arithmetic mean of the mode-wise adiabaticity parameters.

The central result of the present work is not merely the inequality $E_\tau \ge E_0$,
which follows from more general arguments, but rather its explicit mechanical characterization in terms of the generalized Husimi's adiabaticity parameter. Its decomposition into mode-wise adiabaticity parameters reveals how each normal mode contributes to irreversibility in many-body dynamics.

The normal-mode spectrum determines how the network structure affects
irreversibility. In particular, the eigenvalues $\lambda_i$ set the
characteristic response timescales and the contribution of each mode
to the nonadiabatic increase in energy. The present framework therefore
provides a mechanical interpretation of irreversibility in terms of the collective mode structure of the network, 
and its potential applications merit further investigation.

\acknowledgments
This work was supported by JSPS KAKENHI Grant Number 25K07163. During the preparation of this manuscript, the authors used ChatGPT (OpenAI, GPT-5.5) to assist with manuscript editing and to generate the numerical code. The authors reviewed, tested, validated, and critically assessed all AI-assisted outputs and take full responsibility for the content of the manuscript.

\appendix

\section{Proof of the positive definiteness of $A$}
\label{subsec:proof_of_positive_definite}
Let $\bm{z}\neq\bm{0}$ be an arbitrary real vector and define
$\bm{y}=M^{-1/2}\bm{z}$. From Eq.~(\ref{eq:definition_of_A}),
\begin{align}
    \bm{z}^{\top}A\bm{z}
    &=
    \bm{y}^{\top}(L+B)\bm{y} \nonumber\\
    &=
    b_1y_1^2
    +
    \sum_{(i,j)\in E}a_{ij}(y_i-y_j)^2
    \ge 0.
\end{align}
If this quadratic form vanishes, the wall term with $b_1>0$ implies
$y_1=0$, while every spring term implies $y_i=y_j$ for each
$(i,j)\in E$. Because the network is connected, these conditions give
$y_i=0$ for every $i$. Thus, $\bm{y}=\bm{0}$ and hence
$\bm{z}=\bm{0}$, contradicting the assumption. Therefore,
$\bm{z}^{\top}A\bm{z}>0$ for every nonzero $\bm{z}$, so $A$ is
positive definite and all of its eigenvalues are positive.

\section{Derivation of Eq.~(\ref{eq:E_t_N-body})}
\label{app:derivation_of_E_t}

Because the initial Hamiltonian can be decomposed into a sum of independent mode Hamiltonians, the canonical distribution factorizes over the normal
modes. Its second moments are obtained as
\begin{equation}
    \left\langle \left(u_0^{(i)}\right)^2 \right\rangle
    =
    \frac{1}{\beta\lambda_i\omega_0^2},
    \qquad
    \left\langle \left(v_0^{(i)}\right)^2 \right\rangle
    =
    \frac{1}{\beta},
\end{equation}
and
\begin{equation}
    \left\langle u_0^{(i)}v_0^{(i)} \right\rangle=0.
\end{equation}

Substituting Eqs.~(\ref{eq:ut_time_evolution}) and
(\ref{eq:vt_time_evolution}) into the mode Hamiltonian $H_t^{(i)}$ in Eq.~(\ref{eq:decomposition_mode_H}), we can write
\begin{equation}
    H_t^{(i)}
    =
    \alpha_t^{(i)}\left(u_0^{(i)}\right)^2
    +
    \beta_t^{(i)}\left(v_0^{(i)}\right)^2
    +
    \gamma_t^{(i)}u_0^{(i)}v_0^{(i)},
\end{equation}
where
\begin{align}
    \alpha_t^{(i)}
    &=
    \frac{1}{2}
    \left[
    \left(\dot{Y}_t^{(i)}\right)^2
    +
    \lambda_i\omega_t^2\left(Y_t^{(i)}\right)^2
    \right], \\
    \beta_t^{(i)}
    &=
    \frac{1}{2}
    \left[
    \left(\dot{X}_t^{(i)}\right)^2
    +
    \lambda_i\omega_t^2\left(X_t^{(i)}\right)^2
    \right], \\
    \gamma_t^{(i)}
    &=
    \dot{X}_t^{(i)}\dot{Y}_t^{(i)}
    +
    \lambda_i\omega_t^2X_t^{(i)}Y_t^{(i)}.
\end{align}
It follows that
\begin{align}
    \left\langle H_t^{(i)}\right\rangle
    &=
    \frac{1}{\beta}
    \left(
    \frac{\alpha_t^{(i)}}{\lambda_i\omega_0^2}
    +
    \beta_t^{(i)}
    \right) \nonumber\\
    &=
    \frac{1}{\beta}
    \frac{\omega_t}{\omega_0}
    Q_t^{*(i)},
\end{align}
where the second equality follows from
Eq.~(\ref{eq:Husimi's_parameter_N-body}). Summing over all modes and
using $E_0=N/\beta$, we obtain
\begin{align}
    E_t
    &=
    \sum_{i=1}^{N}\left\langle H_t^{(i)}\right\rangle \nonumber\\
    &=
    E_0
    \frac{\omega_t}{\omega_0}
    \left(
    \frac{1}{N}\sum_{i=1}^{N}Q_t^{*(i)}
    \right) \nonumber\\
    &=
    E_0
    \frac{\omega_t}{\omega_0}
    \overline{Q}_t^{*},
\end{align}
which proves Eq.~(\ref{eq:E_t_N-body}).

\section{Derivation of Eq.~(\ref{eq:inequality_for_Q_t^{(i)}})}
\label{app:derivation_of_Q>1_N_body}

For the $i$-th mode, we define the Wronskian
\begin{equation}
    W_t^{(i)}
    \equiv
    \dot{X}_t^{(i)}Y_t^{(i)}
    -
    X_t^{(i)}\dot{Y}_t^{(i)}.
\end{equation}
Because $X_t^{(i)}$ and $Y_t^{(i)}$ both satisfy
Eq.~(\ref{eq:EOM_of_Q_t^(i)}), we obtain
\begin{equation}
    \dot{W}_t^{(i)}
    =
    \ddot{X}_t^{(i)}Y_t^{(i)}
    -
    X_t^{(i)}\ddot{Y}_t^{(i)}
    =0.
\end{equation}
Their initial conditions give $W_0^{(i)}=1$, and hence
$W_t^{(i)}=1$ for all times. Using this identity in
Eq.~(\ref{eq:Husimi's_parameter_N-body}) gives
\begin{align}
    Q_t^{*(i)}-1
    &=
    \frac{\omega_0}{2\omega_t}
    \left(
    \dot{X}_t^{(i)}
    -
    \frac{\omega_t}{\omega_0}Y_t^{(i)}
    \right)^2
    \nonumber\\
    &\quad+
    \frac{\omega_0}{2\omega_t}
    \left(
    \frac{\dot{Y}_t^{(i)}}{\sqrt{\lambda_i}\omega_0}
    +
    \sqrt{\lambda_i}\omega_tX_t^{(i)}
    \right)^2
    \ge0.
\end{align}
Thus, $Q_t^{*(i)}\ge1$, which proves
Eq.~(\ref{eq:inequality_for_Q_t^{(i)}}).

\section{Extension to the wall-free system}
\label{app:no_wall_case}

Setting $B=0$ in Eq.~(\ref{eq:matrix_M_B}), the corresponding mass-weighted coupling matrix $A$ in Eq.~(\ref{eq:definition_of_A}) is altered as
\begin{equation}
    A_{\mathrm{wf}}
    \equiv
    M^{-1/2}LM^{-1/2}.
\end{equation}
This matrix is positive semidefinite. Because the network is connected,
it has a single zero eigenvalue associated with uniform translation:
\begin{equation}
    0
    =
    \lambda_1
    <
    \lambda_2
    \le
    \cdots
    \le
    \lambda_N.
\end{equation}
Writing $M_{\mathrm{tot}}\equiv\sum_{i=1}^{N}m_i$, the normalized
zero-eigenvalue eigenvector can be chosen as
\begin{equation}
    \bm{s}_1
    =
    \frac{M^{1/2}\bm{1}}{\sqrt{M_{\mathrm{tot}}}}.
\end{equation}
The corresponding normal variables are given by
\begin{equation}
    u_t^{(1)}
    =
    \bm{s}_1^{\top}\bm{x}'_t
    =
    \sqrt{M_{\mathrm{tot}}}\,R_{\mathrm{cm},t},
\end{equation}
\begin{equation}
    v_t^{(1)}
    =
    \bm{s}_1^{\top}\bm{p}'_t
    =
    \frac{P_{\mathrm{tot},t}}{\sqrt{M_{\mathrm{tot}}}},
\end{equation}
where $R_{\mathrm{cm},t}\equiv
    \frac{1}{M_{\mathrm{tot}}}
    \sum_{i=1}^{N}m_ix_t^{(i)}$
and $P_{\mathrm{tot},t}\equiv\sum_{i=1}^{N}p_t^{(i)}$
are the center-of-mass position and total momentum, respectively.
The mode equations are obtained as
\begin{equation}
    \ddot{u}_t^{(1)}=0,
    \qquad
    \ddot{u}_t^{(i)}
    =
    -
    \lambda_i \omega_t^2 u_t^{(i)}
    \quad
    (i=2,\dots,N)
\end{equation}
with the corresponding Hamiltonian:
\begin{equation}
    H_t
    =
    \frac{1}{2}
    \left(v_t^{(1)}\right)^2
    +
    \sum_{i=2}^{N}
    \left[
    \frac{1}{2}
    \left(v_t^{(i)}\right)^2
    +
    \frac{\lambda_i\omega_t^2}{2}
    \left(u_t^{(i)}\right)^2
    \right].
\end{equation}
The corresponding partition function is calculated as
\begin{align}
    Z_{\mathrm{wf}}
    &\equiv
    \int_{-\infty}^\infty dv_0^{(1)}
    \prod_{i=2}^{N}
    \left(
    \int_{-\infty}^\infty du_0^{(i)}\int_{-\infty}^\infty dv_0^{(i)}
    \right)
    e^{-\beta H_0}
    \nonumber\\
    &=
    \frac{1}{
    \omega_0^{N-1}
    \sqrt{\lambda_2\lambda_3\cdots\lambda_N}}
    \left(
    \frac{2\pi}{\beta}
    \right)^{N-\frac{1}{2}}.
\end{align}
From this partition function, we can obtain the initial average energy as
\begin{equation}
    E_0
    =
    \langle H_0 \rangle
    =
    -\frac{\partial}{\partial \beta}\ln Z_{\mathrm{wf}}
    =
    \left(
    N-\frac{1}{2}
    \right)
    \frac{1}{\beta}.
\end{equation}

As the kinetic energy of the translational mode is conserved, its average remains
at $1/(2\beta)$. Applying the result of Appendix~\ref{app:derivation_of_E_t}
to the remaining $N-1$ internal modes gives
\begin{align} \label{eq:E_t_N-body_nowall}
    E_t
    &=
    \frac{1}{2\beta}
    +
    \frac{1}{\beta}
    \frac{\omega_t}{\omega_0}
    \sum_{i=2}^{N}Q_t^{*(i)} \nonumber\\
    &=
    E_0
    \frac{\omega_t}{\omega_0}
    \mathcal{Q}_{t,\mathrm{wf}}^{*},
\end{align}
where we have introduced the effective wall-free adiabaticity parameter
\begin{equation}
    \mathcal{Q}_{t,\mathrm{wf}}^{*}
    \equiv
    \frac{1}{N-\frac{1}{2}}
    \left(
    \frac{\omega_0}{2\omega_t}
    +
    \sum_{i=2}^{N}Q_t^{*(i)}
    \right).
\end{equation}
Unlike the generalized Husimi's adiabaticity parameter
$\overline{Q}_t^*$ in Eq.~(\ref{eq:generalized_Husimi_parameter}) for the wall-attached system, this parameter is not a
simple arithmetic mean of mode-wise adiabaticity parameters because it also
contains the contribution from the translational mode.

For a cyclic operation satisfying $\omega_\tau=\omega_0$, the bound
$Q_\tau^{*(i)}\ge1$ gives
\begin{align} \label{eq:Planck_principle_N-body_nowall}
    E_\tau
    &=
    E_0
    \mathcal{Q}_{\tau,\mathrm{wf}}^{*}
    \ge
    E_0.
\end{align}
Thus, even in the wall-free system, the energy non-decrease can be expressed
in terms of the mode-wise adiabaticity parameters together with
the contribution from the translational mode.

\sloppy
\hbadness=10000
\bibliography{reference}

\end{document}